\documentclass[aps,prl,twocolumn,showpacs,superscriptaddress,groupedaddress]{revtex4}  
\usepackage{graphicx}  
\usepackage{dcolumn}   
\usepackage{bm}        
\usepackage{amssymb}   
\begin{document}

\widetext

\title{Broad band turbulent spectra in gamma-ray burst light curves}

\author{Maurice H.P.M. van Putten}
\affiliation{Astronomy and Space Science, Sejong University, 98 Gunja-Dong Gwangin-gu, Seoul 143-747, Korea\\}
\author{Cristiano Guidorzi}
\affiliation{Department of Physics and Earth Sciences, University of Ferrara, via Saragat 1, I-44122, Ferrara, Italy\\}
\author{Filippo Frontera}
\affiliation{Department of Physics and Earth Sciences, University of Ferrara, via Saragat 1, I-44122, Ferrara, Italy, 
and INAF, IASF, Via Gobetti, 101, I-40129 Bologna, Italy}

\date{\today}

%1
\begin{abstract}
Broad band power density spectra offer a window to understanding turbulent behavior in the emission mechanism and, at the highest frequencies, in the putative inner engines powering long GRBs. We describe a chirp search method which steps aside Fourier 
analysis for signal detection in the Poisson noise-dominated 2 kHz sampled BeppoSAX light curves. An efficient numerical 
implementation is described in $O(Nn\log n)$ operations, where $N$ is the number of chirp templates and $n$ is the length of the light
curve time series, suited for embarrassingly parallel processing. For detection of individual chirps of duration $\tau=1$ s, the method is 
one order of magnitude more sensitive in SNR than Fourier analysis. The Fourier-chirp spectra of GRB 010408 and GRB 970816 show a continuation of the spectral slope up to 1 kHz of turbulence identified in low frequency Fourier analysis. The same continuation is
observed in an ensemble averaged spectrum of 40 bright long GRBs. 
An outlook on a similar analysis of upcoming gravitational wave data is included. 
\end{abstract}

\maketitle

\section{1. Introduction}

High frequency power density spectra up to 1 kHz offers a window to the inner engines of long GRBs, that may harbor a rapidly rotating
(proto-)neutron star (PNS) or a black hole-accretion disk or a torus system (BHS; e.g. \cite{pir99,pir04}). Such high frequency window 
may be probed by analyzing the 2 kHz light curves in the BeppoSAX catalogue \cite{fro09} or the upcoming strain amplitude data 
from gravitational wave detectors LIGO-Virgo, KAGRA and the Einstein Telescope \cite{bar99,arc04,kur10,hil08}. 

Given the limited collector area of today's gamma-ray satellites, high frequency light curves of GRBs are typically Poisson noise
dominated, flattening their Fourier spectra above at most tens of Hz. Low-frequency Fourier analysis reveals a Kolmogorov 
spectrum in GRB light curves \cite{bel98,bel00,gui12,dic13}, that is expected to continue smoothly to high frequencies. A broad-band 
turbulent spectrum from the gamma-ray emission process may hereby present a new baseline in searches for high frequency 
modulations by the central PNS or BHS.

%2
Here, we describe a method to extend the turbulent spectra to high frequencies in the Poisson noise of BeppoSAX light curves of 
long GRBs. Turbulence produces phase-coherent intermittencies on short to intermediate time scales, that covers an extended bandwidth in frequency space. To search for turbulence in a Poisson noise dominated signal, we therefore set out to step aside 
Fourier analysis, since it focuses on phase coherence across a narrow bandwidth. We shall apply matched filtering with chirps 
whose frequencies increase or decrease exponentially in time, representing phase coherence across a finite bandwidth.

The computational effort is $O(N n\log n)$ operations, consistent with the Fast Fourier Transform, where $n$ denotes the number of samples in the data time series and $N$ the number of chirp templates. This limit appears to be competitive to other approaches \cite{oto10,fis11}.

To illustrate our method, we report on spectra of bright 2 kHz sampled light curves of long GRBs in the BeppoSAX catalogue \cite{fro09}.

The chirp templates are described in \S2 and an efficient numerical implementation of the matched filtering algorithm is given in \S3. 
Chirp detection sensitivity is analyzed in \S4. We apply the method to extract broad band spectra from a sample of BeppoSAX light curves in \S5,6, to derive a continuation of Fourier spectra. An outlook on further applications is briefly described in \S7.

\section{2. Chirp templates}

%3
Chirps are transients described by a base frequency and a frequency rate of change. They are different from quasi-periodic oscillations (QPOs), that pertain to frequencies meandering about a steady mean. QPOs are notoriously absent in GRB light curves \cite{cen10}, whence they will not be considered here.

%4
Given the limitations of Fourier analysis to extract broad band turbulent spectra in noisy data, we here consider a search by matched filtering for time coherent features across a finite frequency bandwidth. Of particular interest are chirps, here
with an exponential change in frequency as a function of time. They can be extracted by time slicing a single long duration chirp \cite{van11} into  subintervals of duration $\tau$ much shorter than the duration $T$ of a GRB. Starting from frequency $f_0$ and decaying to a late time, asymptotic frequency $f_1$, the frequency evolution is essentially exponential in time,
\begin{eqnarray}
\begin{array}{l}
f(t)=  \,f_1+(f_0-f_1)e^{-\,a t/T}, 
\end{array}
\label{EQN_f}
\end{eqnarray}
where $a$ is a dimensionless scale to parametrize the time scale $T/a$ of change in chirp frequency. 

The choice of $\tau$ is used to search for phase coherence over a time scale $\tau$, such that $\Delta t<< \tau << T$, 
where $\Delta t$ denotes the sampling time interval or the bin size of integration of the data. Accordingly, we consider 
$N=T/\tau$ time intervals
\begin{eqnarray}
t_k < t < t_{k+1},~t_k=\frac{kT}{N}~(k=0,1,\cdots N-1)
\label{EQN_s}
\end{eqnarray}
in our time-slicing procedure. Fig. 1 illustrates slicing of a model chirp with $T=8$ s into $N=8$ one second chirp templates
\begin{eqnarray}
z_k(t^\prime),~0\le t^\prime \le \tau,~t^\prime=t-t_k,
\label{EQN_f2}
\end{eqnarray}
where $k=0,1,\cdots N-1$.

%5
In steady state, the statistical properties of, e.g., a velocity field in turbulent motion are the same when viewed forwards and backwards in time, whereby the probabilities for detecting positive and negative chirps are similar for intermediate durations $\tau$.
We shall therefore employ difference chirps, obtained as the difference of chirps considered forwards and backwards in time. 
If $x_k(t^\prime)$ is a chirp template extracted from (\ref{EQN_f1}), we 
consider
\begin{eqnarray}
x_k(t^\prime) = z_k(t^\prime)-z_k(\tau-t^\prime) ~(0\le t^\prime\le \tau).
\label{EQN_f3}
\end{eqnarray}
%6
In using (\ref{EQN_f3}), we save a factor of two in computational cost, allowing for detections of positive or negative chirps in one calculation. Because the cross-correlation between chirps forward and backwards in time is small, (\ref{EQN_f3}) can be used efficiently with negligible loss in sensitivity over performing matched filtering twice, using positive and negative chirps separately in each run.

\begin{figure}[h]
\centerline{\includegraphics[width=75mm,height=75mm]{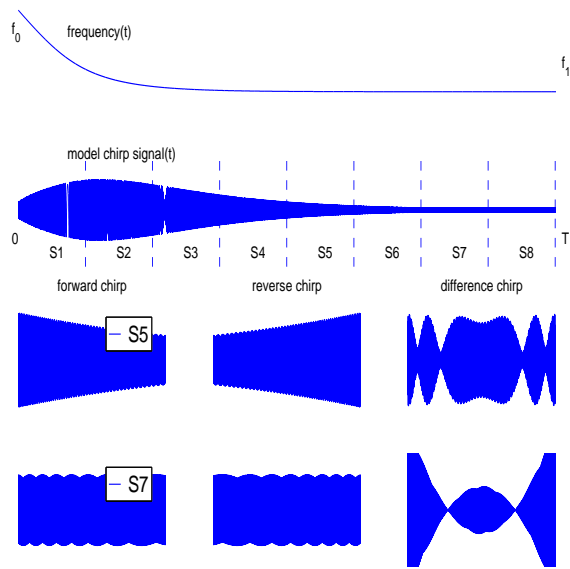}} 
\caption{Shown is a long duration exponential decay in frequency $f$ to $f_1$ ($t=T$) from $f_0$ ($t=0$) of
a model signal ({\em top windows}). Chirps of intermediate duration $\tau=1$ s are extracted by time slicing.
For a chirp search with slews of either sign in matched filtering, difference chirps are used ({\em right}).}
\label{fig1}
\end{figure}

\section{3. Efficient matched filtering}

%7
To develop a search with maximal sensitivity to features with frequencies changing in time, we set out to employ matched filtering.
Matched filtering obtains the highest possible sensitivity for phase-coherent features, upon including sufficiently many templates
to cover the full range of possible signals and their phase-coherent behavior. 
For this reason, efficient numerical implementation of method matched filtering is important.

For a time series $y(t)$ and chirp template $x(t)$, let $\bar{x}(t)=x(t)-\mu_x$ and 
$\bar{y}(t)=y(t)-\mu_y$ by subtracting the respective mean values $\mu_x$ and $\mu_y$, 
and consider their cross correlation
\begin{eqnarray}
\rho(t) = \int_{-\infty}^\infty \bar{x}(s)\bar{y}(t+s) ds.
\label{EQN_f1}
\end{eqnarray}
Using the Fourier transform $F(k)$ of a function $f(t)$,
\begin{eqnarray}
F(k)=\frac{1}{\sqrt{2\pi}} \int_{-\infty}^{\infty} f(t)e^{-ikt} dt,\\ f(t)=\frac{1}{\sqrt{2\pi}} \int_{-\infty}^{\infty} F(k)e^{ikt} dk,
\label{EQN_f2}
\end{eqnarray}
$\rho(t)$ is efficiently calculated as the inverse Fourier transform of the product $X^*(k)Y(k)$ of the complex conjugate $X^*(k)$
of the transform of $\bar{x}(t)$ and $Y(k)$ of $\bar{y}(t)$.

In matched filtering, the potential significance of a chirp is identified by the normalized cross correlation 
$\hat{\rho}(t)$ of $\hat{x}$ and $\hat{y})(t)$ (i.e., the Pearson coefficient)
\begin{eqnarray}
\hat{x}(t)=\frac{\bar{x}(t)}{||\bar{x}||},~\hat{y}(t)=\frac{\bar{y}(t)}{||\bar{y}||},
\label{EQN_f3}
\end{eqnarray}
where $||f(t)||=\{\int_{-\infty}^\infty f^2(t)dt\}^{1/2}$ is the $L^2$ norm of $f(t)$. 

For discrete series of samples at $t=t_i$ $(i=1,\cdots n)$, consider $Y=\{\bar{y}_i\}_{i=1}^n$ and $X=\{{\bar x}_i\}_{i=1}^m$ $(1\le m \le n)$,
where $N=n/m$ denotes the number of slices illustrated in Fig. 1. 
Since (\ref{EQN_f1}) is bilinear, the sample correlation coefficient (SCC) $\rho_i=\rho(t_i)$ obtained from on $Y$ and $X$ is readily calculated by the Fast Fourier Transform (FFT) in $O(n\log n)$ operations. However, the normalized sample correlation coefficient $\hat{\rho}_i$ obtained from 
$\hat{Y}=\{\bar{y}_i\}_{i=1}^n/|| \bar{y}||$ and 
$\hat{X}=\{\bar{x}_i\}_{i=1}^m/||\bar{x}||$ $(1\le m \le n)$
is nonlinear in $Y$ and $X$ due to normalizations $||f||=\{\Sigma f_i^2\}^{1/2}$. Direct evaluation of these
nonlinear expressions is prohibitively expensive when the number of chirps is large.

To be precise, consider the Pearson coefficient between $Y$ given by a section $y_{il}=y_{i+l}$ $(1\le l\le m, 1\le m \le n-m$), 
of the time series $\{y_i\}_{i=1}^n$ and a template $X$,
\begin{eqnarray}
\hat{\rho}_i =  \Sigma_{j=1}^m \hat{x}_j \hat{y}_{ij}~(i=1,2,\cdots,n-m)
\label{EQN_f4}
\end{eqnarray}
where
\begin{eqnarray}
\begin{array}{ll}
\hat{x}_j = \bar{x}_j/ \sqrt{\Sigma_{l=1}^m \bar{x}_l^2},  & \bar{x}_l = x_l - m^{-1}\Sigma_{j=1}^m x_j\\
\hat{y}_{ij} = \bar{y}_{ij}/ \sqrt{\Sigma_{l=1}^m \bar{y}_{il}^2}, & \bar{y}_{il} = y_l - m^{-1}\Sigma_{j=1}^m y_{ij}.
\end{array}
\label{EQN_f5}
\end{eqnarray}
Exact normalization in (\ref{EQN_f4}) ensures it to be the cosine between $Y$ and $X$, satisfying $-1 \le \hat{\rho}_i \le 1.$

The unnormalized cross correlation 
\begin{eqnarray}
\rho_i =  \Sigma_{j=1}^m \bar{x}_j \bar{y}_{ij}~(i=1,2,\cdots,n-m)
\label{EQN_f7}
\end{eqnarray}
can be evaluated in $O(n\log n)$ operations using the Fast Fourier Transform (FFT), which is
is cost effective compared to (\ref{EQN_f4}) when $m>\log n$. We now
consider $\rho_i$ block wise over data slices of duration $\tau$, i.e., the discretized intervals
\begin{eqnarray}
km\le i \le (k+1)m~(k=0,1,\cdots N-1).
\label{EQN_s2}
\end{eqnarray}
Normalizing $\rho_i$ as an array over each slice requires $N$ normalizations, 
\begin{eqnarray}
C_k=\sqrt{\Sigma_{j=km}^{(k+1)m} \rho_{j}^2}~(k=0,1,\cdots N-1).
\label{EQN_f8a}
\end{eqnarray}
The $C_{k_i}$ define a staircase as a function of $i$, where $k_i m\le i \le (k_i+1)m$, giving a block wise normalized SSC 
\begin{eqnarray}
\hat{\rho}_i =  \frac{\rho_i}{C_{k_i}}.
\label{EQN_f8b}
\end{eqnarray}
Block wise normalization is particularly opportune when $y_i$ shows variations in dispersion no faster than the 
intermediate time scale $\tau$. If so, our block normalized SCC applies at the operational cost of FFT, which is below 
the cost of true SCC's in (\ref{EQN_f4}) whenever $m>\log n$.

When $m$ is relatively large, e.g.,  $2^p$ with $p\ge 8$, $\hat{\rho}_i$ typically shows a near-Gaussian distribution 
by the central limit theorem. In what follows, we shall use a further normalization by the variance of the cross 
correlations (\ref{EQN_f7}) in each time slice, i.e., 
\begin{eqnarray}
{R}_i = \frac{\rho_i}{\sigma_{k_i}},~~\sigma_k=C_k/\sqrt{m-1}.
\label{EQN_f8c}
\end{eqnarray}
$R_i$ hereby differs from $\hat{\rho}_i$ only by a constant factor $\sqrt{m-1}$, whose 
Probability Density Function (PDF) approaches a truncated Gaussian of unit variance. The truncation is a function of both the 
total number of trials $n-m$ and the potential for a chirp being present in the data.

Fig. 2 illustrates the numerical implementation.

\begin{figure}[h]
\centerline{\includegraphics[scale=0.47]{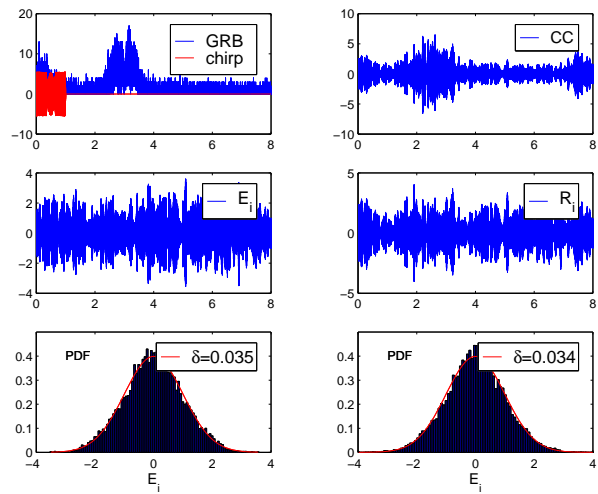}}
\caption{Shown is the light curve of GRB\,010408 and a difference chirp ({\em left top}). The PDF of the moving Pearson 
coefficient $E_i$ ($\hat{\rho}_i$ in (\ref{EQN_f4}) by exact evaluation, normalized to unit variance) 
has a white truncated Gaussian distribution ({\em left}). An efficient numerical 
implementation is obtained by calculating 
the moving cross-correlation CC using FFT 
with subsequent block wise normalization, the PDF of which is also effectively white Gaussian 
({\em right}). Residual deviations $\delta$ with respect to a theoretical Gaussian with unit variance ({\em red}) are a few percent, and 
block wise normalization ({\em right}) yields a PDF on par with the PDF obtained by exact numerical evaluation ({\em left}).}
\label{fig2}
\end{figure}

\section{4. Chirp detection sensitivity}

We consider quantifying the sensitivity of matched filtering relative to that obtained by Fourier analysis. To this end, we perform the 
injection experiment 
\begin{eqnarray}
y(t)=y_0(t) + \alpha y_1(t)
\end{eqnarray}
on the BeppoSAX light curve $y_0(t)$ of GRB\,010408 by the light curve $y_1(t)$ of a chirp for various amplitudes $0.01\le \alpha\le 1$.

A chirp search by matched filtering obtains the time series $R_i$ following (\ref{EQN_f8c}). We consider detection by
\begin{eqnarray}
R = \max_{i=1..n} |R_i|
\label{EQN_R}
\end{eqnarray}
in light of the approximately Gaussian PDF of the $R_i$, truncated by the finite number of $n-m$ trials in each template search. 

To compare (\ref{EQN_R})
with Fourier analysis, we calculate the spectrum by the Welch method \cite{wel67,coo70,pre02} using a partition in $Q=10$ sub-windows of length $n/Q$ 
with a $\Delta f=1$ Hz frequency resolution. The spectrum is calculated as an average over $2Q$ periodograms from intervals of 
length $n/Q$ with 50\% overlap. Each periodogram is obtained with a Welch window, i.e., as the Fourier transform of $w(t)y(t)$ by FFT, 
where $t=4u(1-u)$, where $u=t_1/0.8$ with $t_1/(1~{\rm s})$. 
By construction of the Welch method, fluctuations in the resulting power spectrum are 
approximately Gaussian, as a $\chi^2_{4Q}$ distribution from averaging $2Q$ periodograms. 

For each $\alpha$, detection is expressed, similarly to (\ref{EQN_R}), by peak values 
\begin{eqnarray}
H=\max_{k=1..n/Q} H_k
\label{EQN_W}
\end{eqnarray}
relative to the asymptotically flat Poisson dominated spectrum in terms of
\begin{eqnarray}
H_k=\frac{|c_k| - s_0}{\sigma_0},
\label{EQN_Hk}
\end{eqnarray}
where $s_0$ and $\sigma_0$ denote the mean and standard deviation of the $|c_k|$ in the Poisson dominated tail of the Fourier spectrum.

As a control, consider a chirp search in random data. In this event, $R$ and $H$ have expectation values $R_0$ and $H_0$, 
respectively, that derive from the expectation value $x_0$ of the truncation in the distribution of $n-m$ trials of a variable $x$ taken 
from a Gaussian distribution with unit variance. Here, $x_0$ satisfies
\begin{eqnarray}
N\,\rm{erfc}(x_0/\sqrt2)\simeq 1
\label{EQN_f10}
\end{eqnarray}
with $N=(n-m)$ for $R_0$ and $N=n/Q$, for $H_0$, where $\rm{erfc}(x)=2/\sqrt{\pi}\int_x^\infty e^{-s^2}ds$ denotes the complementary 
error function. 

\begin{figure}[h]
\centerline{\includegraphics[width=80mm,height=40mm]{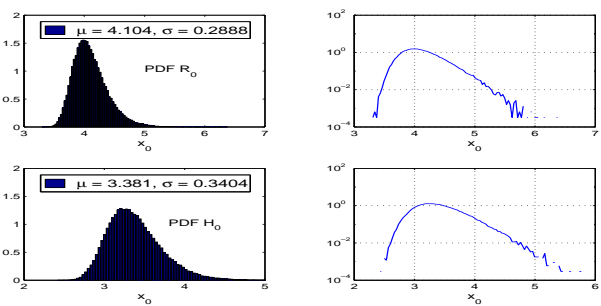}}
\caption{({\em Left.}) Shown is the skewed distribution of $M=10^5$ maxima $R_0$ in (\ref{EQN_R}) in $n=2^{14}$ trials from 
a Gaussian distribution of unit variance. This distribution has positive skewness with an approximately exponential tail at large arguments 
({\em right}). A similar result with lower mean value obtains for the distribution of maxima $H_0$ in (\ref{EQN_Hk}) with $Q=10$.}
\label{fig3}
\end{figure}

Fig. \ref{fig3} shows $R_0$ and $H_0$ distributions computed numerically from $M=10^5$ maxima 
in trial samples of size $N$ from a Gaussian distribution with unit variance. Their mean $\mu$ is determined by $N$ according to
(\ref{EQN_f10}) and the resulting distributions have positive skewness. The distribution shown are themselves truncated 
according to (\ref{EQN_f10}) upon substituting $MN$ for $N$.

Figs. 4 and 5 show the results of our injection experiment for various values of $\alpha$ obtained in Fourier analysis and, respectively,
matched filtering. 

Applied to $R_i$, (\ref{EQN_f10}) implies a base level $R_0\simeq 4$ when $n=2^{14}$ in considering 8 seconds of the 2 kHz light curves in the BeppoSAX catalogue. Under the null hypothesis of no signal present, $R$ has a probability of occurrence $P\simeq (n-m)\,{\rm erfc}(R/\sqrt2)$. An excess $R>R_0$ is a false positive with $P<1$ or denotes the presence of a signal. For $m=2048$ ($\tau=1$ s), a $3\sigma$ detection corresponds to $R_3=5.24$, indicated by the dot-dashed line in Fig. 5. $H$ obtains similarly from $H_k$
giving a $3\sigma$ threshold $H_3=4.79$.

\begin{figure}[h]
\centerline{\includegraphics[scale=0.47]{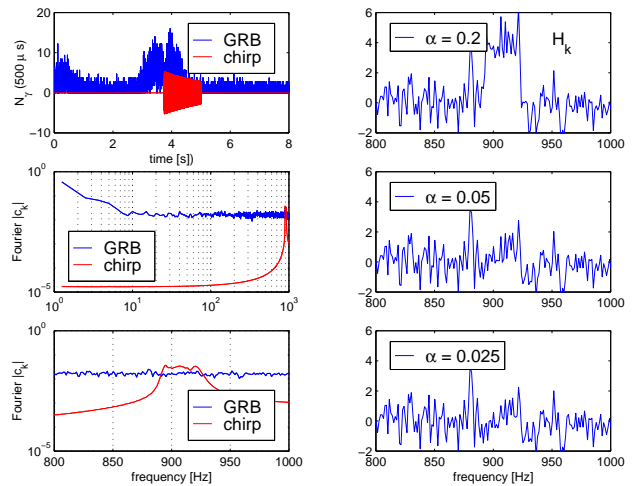}}
\caption{Overview of injection experiments of a chirp template in the 2 kHz sampled BeppoSAX light curve of GRB 010408 ({\em left top}).
The Fourier spectra are obtained by the Welch method. The spectrum of the GRB is asymptotically flat due to Poisson noise, and the spectrum of the chirp is effectively of finite band width ({\em left}). Results of detection by Fourier analysis are shown for various injections parametrized by $\alpha$ ({\em right}).}
\label{fig4}
\end{figure}

\begin{figure}[h]
\centerline{\includegraphics[scale=0.47]{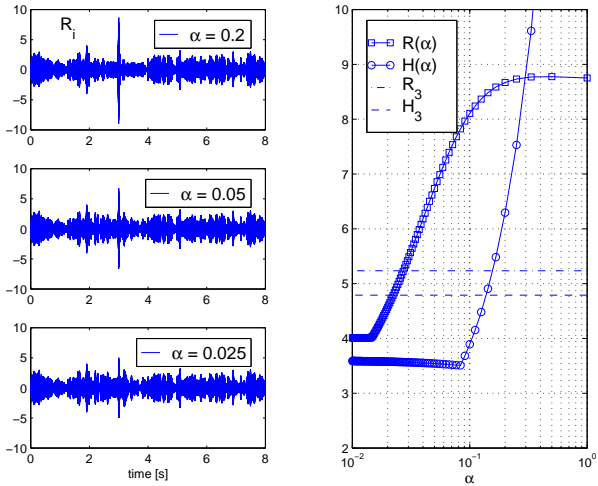}}
\caption{Shown are various results of the injection experiment by matched filtering ({\em left}). The results of 
matched filtering show a gain in sensitivity over Fourier analysis for a $3\sigma$ detection by about one order of magnitude
in SNR=$\alpha^2$ ({\em right}).}
\label{fig5}
\end{figure}

For a $3\sigma$ detection, the critical value $\alpha_3=0.027$ for matched filtering is a factor of 5.3 smaller than $\alpha_3=0.14$ for 
Fourier analysis using the Welch method. An additional moving average of the Fourier spectrum shows some improvement in its
sensitivity, leaving a gain by matched filtering by about one order of magnitude in SNR. 

The sensitivity of matched filtering shown in Fig. 5 is consistent with the theoretical sensitivity limit of $1/\sqrt{m}\simeq 2.2\%$, where 
$m=2048$ in the case at hand for $\tau=1$ s. The observed value $\alpha=2.5\%$ for a 1-$\sigma$ excess in $R$ above background 
is indeed close to the anticipated value $1/\sqrt{m}$. 

\section{5. Chirp spectra $h(f)$}

We observe that both $R(\alpha)$ scales linearly with $\alpha$ across an appreciable range beyond $\alpha_c\simeq 0.015$,
below which it assumes the constant background value $R_0$ set by the number of trials $n-m$. Upon subtracting $R_0$,
we thus obtain a detection method with a linear response to small amplitude signals. In general, $R_0$ is a function of 
frequency, which poses the question on devising a suitable control.

We express the spectra in terms of a strain $h=h(f)$ given by the square root of the PDS, 
\begin{eqnarray}
h(f)=\frac{R(f) - R_0(f)}{R_0(f)\sqrt{B(f)}},~h_k=\frac{|c_k|-s_0}{\sigma_0 \sqrt{B_0}}.
\end{eqnarray}
Here, $R_0(f)$ denotes the results of chirp analysis of control light curves and
$B(f)=\kappa(\tau)f^{1/2}$ is the bandwidth of a chirp templates about the frequency $f$, 
where $\kappa(\tau)$ is a coefficient that depends on the choice of chirp duration. The
$h_k$ are calculated from the Fourier coefficients $|c_k|$, that have a noise dominated high frequency tail with
standard deviation $\sigma_0$ and mean $s_0$.

A chirp search is not a linear transform in the sense of Fourier analysis. A chirp search seeks a best-fit chirp to the data, by varying frequency and frequency rate of change. Different chirp templates are hereby linearly dependent at high resolutions even though they may retain finite cross-correlations, as opposed to working with basis functions that satisfy exact linear independence. However, this distinction is immaterial in calculating spectra.

For Gaussian additive noise, such as in the high frequency, shot-noise dominated region of strain amplitude
noise in gravitational wave detectors, light curves obtained by time randomization or produced by a random number 
generator will be effective as a control $R_0$. By whitening in (\ref{EQN_f8c}), these two alternatives give 
essentially the same results. 

\begin{figure}[h]
\centerline{\includegraphics[scale=0.47]{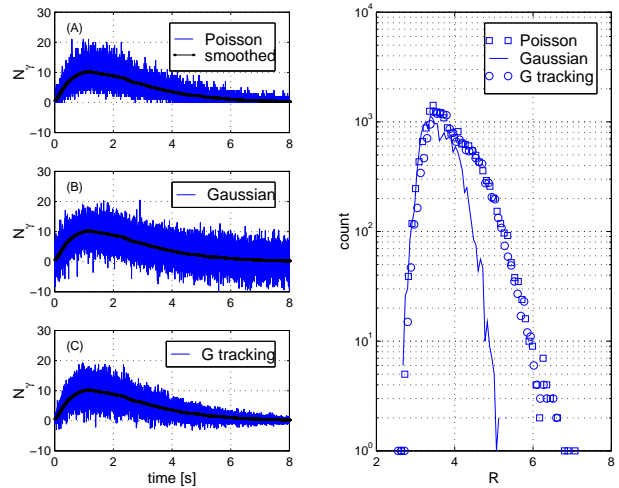}}
\caption{({\em Left}): Top panel (A) shows a synthetic fast rise and exponential decay Poisson noise light curve (FRED) with a 
smoothed light curve following a 2 Hz filter, and derived light curves with (B) Gaussian additive noise and (C) Gaussian noise, 
whose variance tracks the amplitude of the smoothed light curve. ({\em Right}): The $R$ distribution is shown for different noise 
types added to the smoothed light curve, each with positive skewness as in Fig. 3, here obtained by matched filtering over
$N=160000$ chirp templates with log-uniform distribution in frequency and frequency rate of change. 
The distribution of $R$ of (A) shows a pronounced excess to that of (B). Essentially the same $R$ distribution results from (C). 
The excess is therefore due to the Poisson correlation between variance and average, here a moving average
defined by the smoothed light curve.}
\label{fig6}
\end{figure}

For the Poisson noise in the 2 kHz BeppoSAX light curves, whose average photon counts are of the order of unity per 0.5 ms bin, 
we propose as a control $R_0$ a synthetic Poisson noise light curve, that shares the same smoothed light curve as the original.
A control of this type accurately captures the secular variation of the variance in the noise with (smoothed) amplitude, illustrated
in Fig. \ref{fig6}.

We are now in a position to apply our method to two bright long GRBs from the BeppoSAX catalogue. Matched filtering
calculations are performed with a log-unfiorm distribution in frequency and frequency rate of change over a total of 
5.76 million templates. For control, we use synthetic Poisson noise light curves about smoothed light curves following a 
low-pass filter at 2 Hz. 

%B1
Fig. 7 shows a blended Fourier-chirp spectrum up to about the maximal frequency of 1000 Hz set by the sample rate of 2 kHz.
Here, the low frequency spectrum is computed by Fourier analysis and the high frequency spectrum by our
chirp search method. Included is a linear extrapolation of the low frequency spectrum, to highlight a common spectral slope
in the low and high frequency spectra, here obtained independently by two completely different methods.

\begin{figure}[h]
\centerline{\includegraphics[width=90mm,height=55mm]{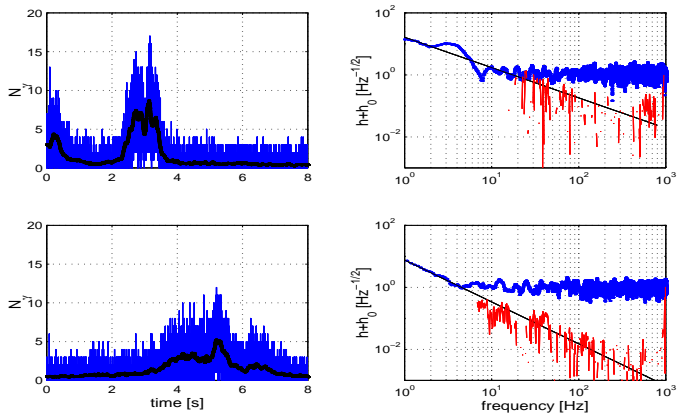}}
\caption{Shown are two bright long GRBs in the BeppoSAX catalogue (GRB 010408 and 970816) 
at 2 kHz sampling over the first 8 seconds, their smoothed light curves ({\em black}) and their spectra over 1-1000 Hz in a log-log plot. 
Fourier analysis reveals a typical low frequency turbulent spectrum, noise limited above at most tens of Hz ({\em blue}) shown with asymptotic normalization $h_0=1$. The spectral slope identified at low frequency in Fourier analysis ({\em black solid lines}) continues at high frequency 
in our matched filtered chirp search, here over 5.76 million templates ({\em red}, $h_0=0$). }
\label{fig7}
\end{figure}

%B2
\section{6. Ensemble spectrum of bright GRBs}

\begin{figure}[h]
\centerline{\includegraphics[width=80mm,height=55mm]{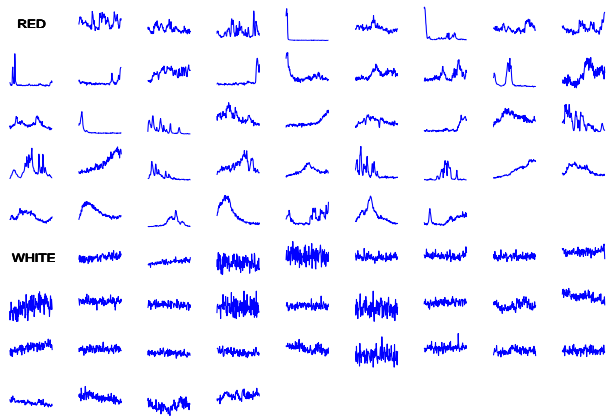}}
\caption{Shown are the smoothed light curves (sorted by $T_{90}=3-456$ s) of an ensemble of 72 bright long GRBs in 
the BeppoSax catalogue, sampled at 2 kHz over the first 8-10 s.}
\label{fig9}
\end{figure}

\begin{figure}[h]
\centerline{\includegraphics[width=80mm,height=55mm]{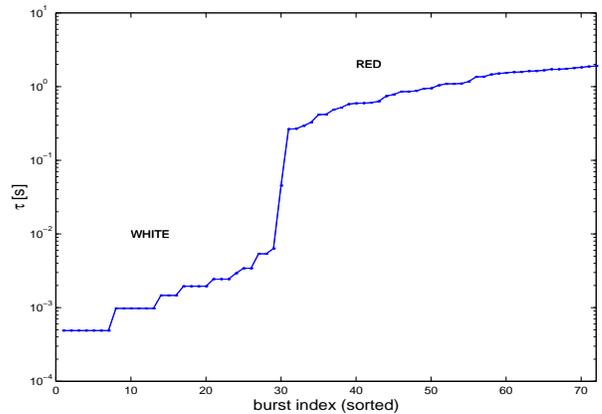}}
\caption{The ensemble of 72 GRBs falls into two groups according to the first zero in their autocorrelation coefficients.
On average, red (40) and white (32) bursts have a first zero at 1.08 s and, respectively, 0.003 s. Their mean
durations $T_{90}$ are, respectively, 48.93 s and 113.2 s.}
\label{fig10}
\end{figure}

We next consider a sample of 72 bright events in the BeppoSAX catalogue (Fig. \ref{fig9}). We select a subsample 
of 40 events with a pronounced autocorrelation in their 2 kHz light curves (``red," Fig. \ref{fig10}) for extracting 
an ensemble averaged Fourier-chirp spectrum.


\begin{figure}[h]
\centerline{\includegraphics[width=80mm,height=55mm]{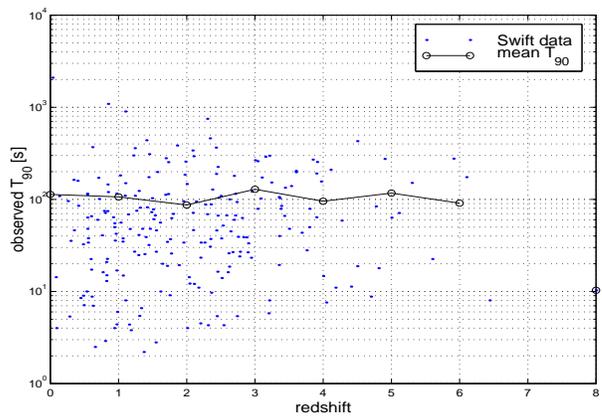}}
\caption{For reference, shown are the observed durations $T_{90}$ versus redshift of long GRBs 
in the {\em Swift} catalogue \cite{swift}. The spread on
observed $T_{90}$ durations is intrinsic, given the lack of correlation to redshift.}
\label{fig8}
\end{figure}

The {\em Swift} catalogue of long GRBs shows no correlation between the observed durations $T_{90}$ 
redshift, shown in Fig. \ref{fig8}. The spread in the observed durations $T_{90}$, therefore, is, essentially intrinsic to the source. 

Based on \cite{van11b}, we consider long GRBs to be produced by black hole-disk or torus systems (BHS), rather than a 
(proto-)neutron star, based on two hyper-energetic GRB-supernovae with an output exceeding the maximal spin energy 
of the latter. In a BHS, $T_{90}$ can be identified with the lifetime of black hole spin, whereby $T_{90}\propto M$ for a
black hole mass $M$ \cite{van01}. The spectrum of turbulence and intermittencies in the surrounding accretion disk or torus scales likewise with $M^{-1}$. If correlated to the wind from the disk or torus, the spectrum of the turbulent outflow is normalizable by multiplication by $M$, i.e., by $T_{90}$. 

\begin{figure}[h]
\centerline{\includegraphics[width=80mm,height=55mm]{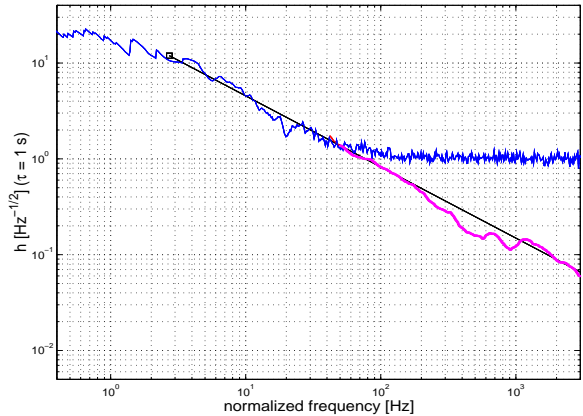}}
\caption{Shown is the continuation of the ensemble averaged spectrum to 1 kHz of 40 bright long GRBs in the BeppoSAX catalogue  
with pronounced autocorrelations. The same spectral slope identified at low frequency in Fourier analysis ({\em black solid line}, $h_0=1$) is found at high frequency in the ensemble average spectrum of 40 bright long GRBs from the BeppoSAX catalogue, 
obtained by a search over 5.76 million chirp templates ({\em purple}, $h_0=0$). The spectrum shown is smoothed over frequency. }
\label{fig11}
\end{figure}

Fig. \ref{fig11} shows the ensemble average of the normalized spectra, plotted as a function of normalized frequency
in the co-moving frame of reference, assuming a fiducial redshift $z=2$ similar to the mean of $z=2.1$ in the {\em Swift}
sample shown in Fig. \ref{fig8}. 

A detailed consideration of alternative chirps shows the following. Chirps with constant amplitude produce slightly more
scatter than those obtained from time slicing shown in Fig. 1. Chirps with $\tau=0.5$ s ($\tau = 2$ s) show considerably
more (less) scatter in the ensemble average. Our choice of $\tau=1$ appears to provide a compromise between scatter
and frequency coverage in extending the slope of the turbulent spectrum.

\section{7. Conclusions}

Turbulent spectra in low frequency in Fourier analysis of long GRBs are found to have a continuation to high frequencies, 
here found in two relatively bright GRBs obtained in a broad band chirp search by matched filtering. 
Matched filtering theoretically obtains maximal sensitivity for a detection, provided that the template bank is sufficiently dense and broad
to capture the signal of interest. To capture turbulence, we here employ chirps with frequencies varying slowly in time following exponential 
decay or growth. Fig. \ref{fig7} shows that Poisson noise is hereby effectively circumvented, upon using a control that shares the secular 
evolution of variance with amplitude of Poisson noise in the 2 kHz BeppoSAX light curves (\ref{fig6}).

Fig. \ref{fig7} shows that extraction of high frequency spectra are quite noisy due to the strong Poisson noise in the BeppoSAX
light curves in light of the small number photon counts in each 0.5 ms bin. On this basis and our limited chirp parameter scan, e.g., 
using $\tau=1$ only, there is no conclusive evidence for the presence or absence of pronounced transient chirps distinct from 
those arising from turbulence. Extensive searches for transient chirps of different durations await a future investigation.

Our extension of the turbulent spectrum in the blended Fourier-chirp spectrum can serve as a new base line in searches for 
high frequency transient features. To this end, we consider the smoothed ensemble averaged spectrum of Fig. \ref{fig9}. 
It may serve as a reference in searches for bumps at high frequency, e.g., around the de-redshifted frequency of 
1 kHz. Detection of a bump would reveal the presence a PNS with misaligned axis of angular momentum and magnetic field, representative for the birth of a new pulsar. The same would be absent in case of rapidly rotating black hole, whose magnetic 
moment and angular momentum are perfectly aligned by Carter's theorem \cite{car68}. Based on the present chirp search 
over 5.76 million templates, no such bump is found.

Figs. \ref{fig7} and \ref{fig11} demonstrate high frequency analysis as a new probe of the physics of the gamma-ray 
emission mechanism, that includes a potentially powerful window to intermittencies in the GRB inner engine even
in light of exceedingly small photon counts. The ensemble of 40 bursts used in 
Fig. \ref{fig11} represents bright events with a pronounced autocorrelation (``red" events with mean photon counts of 1.2569 per bin in the brightest channel) selected out of an initial sample of 72 bright events in the BeppoSAX catalogue, the remaining 32 (``white" events with mean photon counts of 0.5936 per bin in the brightest channel) lacking any perceptible autocorrelation with photon counts 
lower by a factor of about two. Thus, future gamma-ray missions with larger photon yields promise to greatly facilitate high 
frequency analysis, by improving signal-to-noise ratios and enlarging the sample of red events.

Chirp searches can also be applied the strain amplitude data from upcoming advanced gravitational wave
detectors LIGO-Virgo and KAGRA, by changing control to, e.g., time randomized data.
The proposed Fourier-chirp spectra can be extracted to search for gravitational 
wave signatures of possibly forced turbulence 
\cite{van01} in high density matter in the inner disk or torus around black holes, long lasting over up to tens of seconds and 
possibly accompanied by pronounced transient chirps \cite{van11}. Given the limited 
sensitivity range of these detectors, of interest are LGRBs and hyper-energetic core-collapse supernovae in the Local Universe.
Core-collapse supernovae may be found in nearby galaxies such as M51 (hosting SN1994i, SN2005cs, SN 2011dh) and possibly
M82 \cite{mux10} with event rates over one per decade in each.
 
{\bf Acknowledgment.} The BeppoSAX mission was an effort of the Italian Space Agency ASI with participation of 
The Netherlands Space Agency NIVR. Computations in this research were supported in part by the National Science 
Foundation through TeraGrid (now XSEDE) resources provided by Purdue University under grant number TG-DMS100033. 
Some calculations were performed at CAC/KIAS and KISTI. F.~F. and C.~G. acknowledge financial support from Italian Ministry of 
Education, University and Research through the PRIN-MIUR 2009 project on Gamma Ray Bursts (Prot. 2009 ERC3HT).

\end{document}